\documentclass[a4paper]{article}
\usepackage{jheppub}
\usepackage[T1]{fontenc} 
\usepackage[utf8]{inputenc} 
\usepackage{microtype} 
\usepackage{simpler-wick}
\usepackage{mdframed} 
\usepackage{graphicx}
\usepackage{todonotes} 
\usepackage{hyperref}
\usepackage{color}
\usepackage{enumitem}
\definecolor{dark-red}{rgb}{0.4,0.15,0.15}
\definecolor{dark-blue}{rgb}{0.15,0.15,0.4}
\definecolor{medium-blue}{rgb}{0,0,0.5}
\hypersetup{colorlinks, linkcolor={dark-red}, citecolor={dark-blue}, urlcolor={medium-blue}}
\usepackage{amsmath,amsfonts,amssymb}
\usepackage{mleftright} \mleftright
\usepackage{tensor} 
\usepackage{braket} 
\usepackage{mathrsfs} 
\usepackage{leftindex}
\newcommand*{\dd}{\mathop{}\!d}

\newcommand*{\scri}{\ensuremath{\mathscr{I}}}

\newcommand*{\zbar}{\bar{z}}

\newcommand*{\mE}{\mathcal{E}}
\newcommand*{\veps}{\varepsilon}

\newcommand{\Ed}[2]{\mathcal{E}^{[#1]}_{#2}}
\newcommand{\Ehel}[2]{\mathcal{E}^{(#1)}_{#2}}

\DeclareMathOperator*{\SumInt}{%
\mathchoice%
  {\ooalign{$\displaystyle\sum$\cr\hidewidth$\displaystyle\int$\hidewidth\cr}}
  {\ooalign{\raisebox{.14\height}{\scalebox{.7}{$\textstyle\sum$}}\cr\hidewidth$\textstyle\int$\hidewidth\cr}}
  {\ooalign{\raisebox{.2\height}{\scalebox{.6}{$\scriptstyle\sum$}}\cr$\scriptstyle\int$\cr}}
  {\ooalign{\raisebox{.2\height}{\scalebox{.6}{$\scriptstyle\sum$}}\cr$\scriptstyle\int$\cr}}
}
\title{Energy Detectors and Asymptotic Symmetries}
\author{Hern\'an A. Gonz\'alez$^{\; a}$\,\,,}
\emailAdd{hernan.gonzalez@uss.cl}
\affiliation[a]{Facultad de Ingeniería, Universidad San Sebastián,\\ Santiago 8420524, Chile.}
\author{Jakob Salzer$^{\; b}$}
\emailAdd{jakob.salzer@ulb.be}
\affiliation[b]{Université Libre de Bruxelles and International Solvay Institutes,\\
ULB-Campus Plaine CP231, 1050~Brussels, Belgium.}

\abstract{We study detector operators measuring energy to a power $\Delta-2$ at null infinity in four-dimensional gauge theories and gravity. These operators transform as conformal primaries on the celestial sphere and provide a natural basis for describing energy-flux observables in scattering processes. Using the collinear factorization of scattering amplitudes, we derive the universal leading structure of the operator product expansion. 
A key consequence of our analysis is the precise identification of the $\Delta=2$ detector, the number operator. Exploiting the fact that soft charges generate symmetries of the $ S$-matrix, we demonstrate that the number of particles is entirely determined by the product of two soft currents: in gravity, the operator is the square of the supertranslation generator, while in Yang–Mills yields a product of $SU(N)$ Kac-Moody soft currents. This work establishes thus a direct link between detector observables and the soft sector of celestial holography.
}

\makeatletter
\gdef\@fpheader{}
\makeatother
\begin{document}
\maketitle
\section{Introduction}
Massless gauge theories and gravity have a remarkably rich infrared (IR) structure. Every scattering event radiates soft quanta to null infinity, generically obstructing the existence of the conventional S-matrix. Yet, this long-wavelength remnant also carries memory and transforms non-trivially under large asymptotic symmetries~\cite{Bondi:1962px,Sachs:1962wk,Ashtekar:1981bq,Strominger:2013jfa,He:2014laa,He:2015zea,Barnich:2010eb}. Understanding how to properly treat these IR degrees of freedom and to provide a clear-cut definition of well-defined, physically relevant observables in asymptotically flat spacetimes is a fundamental challenge.

Detector operators can provide an interesting perspective on this question \cite{Hofman:2008ar,Chen:2021gdk,Caron-Huot:2022eqs,Caron-Huot:2023vxl,Herrmann:2024yai,Chang:2025zib,Moult:2025nhu}. Placed on the celestial sphere at future null infinity $\mathscr{I}^+$, they register the quantum numbers of particles reaching the null boundary and transform as conformal primaries under the Lorentz group, regarded as the Euclidean conformal group of the celestial sphere \cite{Chang:2022ryc,Chen:2022jhb,Caron-Huot:2022eqs,Hu:2022txx, Hu:2023geb}. Furthermore, they have a very clean theoretical description and are either IR-finite by their very definition or are expected to be IR renormalizable in perturbation theory \cite{Caron-Huot:2022eqs}.

The prime example of a detector operator is the energy operator, given by a null integral of the stress tensor of the outgoing radiation. The same object can also be regarded as the hard contribution to the supertranslation charge that generates an infinite-dimensional symmetry algebra and is conserved in any scattering process \cite{Strominger:2013jfa}. This suggests that, in general, detector observables do not merely measure radiation; they encode the symmetry structure of the infrared.
Building on a small body of work~\cite{Gonzo:2020xza,Hu:2022txx, Hu:2023geb,Herrmann:2024yai}, this paper aims to connect detector operators with asymptotic symmetries that play a crucial role in recent studies of flat-space holography; see \cite{Strominger:2017zoo,McLoughlin:2022ljp,Donnay:2023mrd,Bagchi:2025vri} for reviews on the celestial and Carrollian approach.

In this work, we make progress on three complementary fronts: 
\begin{description}[leftmargin=0pt, labelindent=0pt]
    \item[Detector operator product expansion in quantum gravity.] Using the universal collinear factorization of scattering amplitudes, we show that the leading-order operator product expansion (OPE) of detectors measuring $E^{\Delta-2}$, i.e., energy to some power, closes within a finite set of operators. The expansion automatically generates spinning detectors, which are labeled by representations of ${\rm SO}(2)$. Notably, the coefficients of the OPE are completely fixed by splitting functions at leading order in the coupling constant and are thus universal: they depend on the collinear structure of the theory.
    \item[Connection to holographic celestial OPE.] For self-dual gauge theories, we use the celestial OPE found in~\cite{Pate:2019lpp} to rederive the detector OPE obtained from the splitting function. This computation provides a holographic prescription for the generalized energy operator, defined as the integrated product of celestial primaries over their conformal dimensions.
    \item[Number operator and asymptotic symmetries.] We provide a new physical interpretation of the $\Delta=2$ detector: the operator that counts the number of particles. Using the soft charges that act as symmetries of the $S$-matrix, we show that particle number is encoded in the product of two soft currents: $(i)$ the supertranslation current in gravity and $(ii)$ the ${\rm SU}(N)$ large Kac--Moody current in gauge theory.
\end{description}

We organize the article as follows. In Section~\ref{sec:detectorsdef}, we introduce the boundary fields at $\mathscr{I}^{+}$ and use them to define generalized detector operators as bilocals written in the Carrollian and celestial bases. In Section~\ref{sec:OPE}, we derive the detector OPE at tree level from collinear factorization of scattering amplitudes and show how scalar and spinning detectors appear within a single universal structure. Section~\ref{sec:celestial} reinterprets this expansion from the perspective of celestial amplitudes, highlighting its relation to the celestial operator product expansion in the case of self-dual theories. In Section~\ref{sec:softcor}, we identify the $\Delta = 2$ detector with the particle number operator and show that it can be expressed as a bilinear in soft symmetry currents in both gauge theory and gravity. In Section~\ref{sec:conclusion} we provide some concluding remarks. We give a short summary of our conventions in Appendix~\ref{sec:conventions}.

\section{Definition of generalized energy operators}
\label{sec:detectorsdef}
Detector operators are asymptotic flux operators placed at null infinity, designed to measure the energy or other conserved quantities carried by radiation in specific directions on the celestial sphere. Originally studied in QCD \cite{Korchemsky:1999kt,Sveshnikov:1995vi}, where energy flow is a standard IR and
collinear-safe observable (see the review \cite{Moult:2025nhu}), these observables have since been intimately connected to light-ray or detector operators, extensively developed within Lorentzian CFTs. This connection was sparked by the seminal work \cite{Hofman:2008ar} and subsequently refined in, e.g., \cite{Caron-Huot:2022eqs,Kologlu:2019mfz,Chang:2020qpj}.

Since detector operators measure properties of the radiation flux, we can formulate these fields directly on future null infinity in terms of the Carrollian primary fields $\phi^{(s)}_{J}(u,x^i)$, defined intrinsically on $\scri$, which encode the radiative degrees of freedom of fields in the bulk. We will review their definition briefly in the following.
\subsection{Building detector operators from Carrollian fields}
\label{sec:Carrolldetector}
Outgoing radiation in gauge theory and gravity in asymptotically flat spacetimes admits a natural description at future null infinity $\mathscr{I}^+$, where fields are characterized by the retarded time $u$ and the direction $x^{i}$ on the celestial sphere, as well as any other quantum number such as spin and charge. 
This can be extracted from the bulk fields by imposing a suitable gauge, rescaling with appropriate powers of the radial coordinate and pushing the fields to the boundary $\mathscr{I}$. Performing this procedure for a bulk field of spin $s$ in a $d+2$-dimensional spacetime, one obtains a boundary field $\phi^{(s)}_J(u,x^i)$ with $J=\{j_1,\ldots,j_s\}$ being a multi-index for an irreducible representation of $\mathfrak{so}(d)$; see the summary in Appendix~\ref{sec:bulktobdy} and the discussion in \cite{Donnay:2022wvx} for more details. The field can also carry additional labels in case the bulk field carries some internal symmetry.

These fields can be understood as being defined intrinsically on the boundary in the following way. Null infinity of any asymptotically flat spacetime has the structure of a homogeneous space of the Poincaré group~\cite{Herfray:2021qmp,Figueroa-OFarrill:2021sxz}. Inducing a representation of the Poincaré group from the stabilizer subgroup of this homogeneous space, one lands precisely on the boundary fields defined above~\cite{Banerjee:2018gce,Nguyen:2023vfz}. Since the homogeneous space comes equipped with a \emph{conformal Carrollian structure}, i.e., a degenerate metric and a preferred vector field defined up to a conformal factor~\cite{Geroch:1977jn,Ashtekar:1981bq}, these fields are called \emph{Carrollian primary fields}; see \eqref{Carrollian induced rep} for the defining representation. This provides the starting point for the program of Carrollian holography~\cite{Donnay:2022aba,Bagchi:2022emh,Donnay:2022wvx}.

For instance, in the case of gravity one finds the gravitational shear $C_{ij}$, characterizing outgoing radiation in Bondi coordinates, as being given by the boundary field $\phi^{(2)}_{ij}(u,x^i)$ that is traceless and symmetric. Similarly, the field $\phi^{(1)a}_i(u,x^i)$ gives the leading order in the spatial components of a bulk gauge field $A^{a}_\mu(x^i)$ that carries the propagating degrees of freedom.

For our purposes, these boundary fields will serve as convenient building blocks for constructing local and bilocal operators at null infinity. We refer the reader to \cite{Banerjee:2018gce,Donnay:2022wvx,Nguyen:2023vfz} for more details.
\subsection{The generalized energy operator}
\label{sec:genenergy}
Using the above building blocks, one can start building more complicated objects by taking composites of Carrollian primary fields and integrate them against appropriate kernels. Although one could easily imagine more general objects, the main observable of interest for our purposes are operators that measure a power of the outgoing energy. These are given by
\begin{equation}
  \label{eq:detectordef}
  \Ed{s}{\Delta}(x^i)=\mathcal{C}_{\Delta,s}\int\dd u_1 \dd u_2\frac{\mathbb{P}^{[s]}[\phi^{(s_1)}(u_1,x^i)\phi^{(s_1)}(u_2,x^i)]}{(u_1-u_2- i \epsilon)^{a}}.
\end{equation}
Here, we choose an $i\epsilon$ prescription to regulate the pole and $\mathcal{C}_{\Delta,s}$ is a normalization constant. 
Setting 
\begin{equation}
  \label{eq:weightfix}
  \Delta=a-2+d.
\end{equation}
and furthermore projecting to a spin-$s$ component of the tensor product representation $s_1\otimes s_1$ denoted by $\mathbb{P}^{[s]}$, the field $\Ed{s}{\Delta}(x^i)$ transforms as a primary field of weight $\Delta$ and spin $s$ under the $d$-dimensional Euclidean conformal group, i.e., the Lorentz group in $d+2$ dimensions.\footnote{Here, we are following the conventions of the celestial and Carrollian CFT literature and regard $\Ed{s}{\Delta}$ as an operator of a fiducial two-dimensional CFT on the boundary labelled by $(\Delta,s)$, where $\Delta$ is the eigenvalue of a Lorentz boost in the direction of the north pole of the celestial sphere. In most of the literature on detector operators, the operator is instead labelled by a continuous Lorentz spin $J^{\textrm{there}}_L=-\Delta^{\textrm{here}}$ with $s$ being called the transverse spin. This perspective is more natural when regarding detector operators as light transforms of local operators in a Lorentzian CFT \cite{Kravchuk:2018htv}. } Equivalently, $\Ed{s}{\Delta}(x^i)$ transforms as a zero-momentum Carrollian primary field $\partial_{u}\Ed{s}{\Delta}(x^i)=0$. The form of the integral kernel is fixed by these requirements.
If the field $\phi^{(s)}_J$ has any additional labels such as color, one would also naturally include a sum over these labels.

The simplest choice is the scalar ``trace'' representation $s=0$. We then have
\begin{equation}
  \label{eq:detector}
\Ed{0}{\Delta}\equiv\mathcal{C}_{\Delta}\int\dd u_1 \dd u_2\frac{\sum_{J}\phi^{(s)}_{J}(u_1,x^i)\,\phi^{(s)}_{J}(u_2,x^i)}{(u_1-u_2- i \epsilon)^{\Delta+2-d}}
\end{equation}
where the sum goes over all polarizations. This operator measures the null-integrated energy density on the sphere if $\Delta=d+1$. For generic $\Delta$, the formula above yields the generalized energy operator (GEO)\footnote{In the context of QCD, these are known as Dokshitzer--Gribov--Lipatov--Altarelli--Parisi (DGLAP) operators \cite{Chang:2025zib}. For $\Delta=d+1$ this is usually known as the ANEC operator.} defined in \cite{Kravchuk:2018htv,Caron-Huot:2022eqs,Chen:2021gdk}.

While the generalized energy detector defined above is certainly the most important example of an operator of the form \eqref{eq:detectordef}, we will find that the operator product expansions discussed below also naturally lead to detector operators with intrinsic spin.

Let us focus, for concreteness, on the case in which the underlying Carrollian field has spin $s=1$, describing a gauge field in the bulk. The tensor product of two vectors decomposes into a scalar representation, an antisymmetric two-tensor, and a symmetric traceless two-tensor. The scalar representation gives rise to the GEO $\Ed{0}{\Delta}$ defined in \eqref{eq:detector}, but from the general formula \eqref{eq:detectordef} we can also define detectors that transform as spin two and two-form (or adjoint) representation of $\mathfrak{so}(d)$, respectively.
These detectors measure again energy-to-some-power but are now sensitive to the energy contained in different polarizations.
In the case where both operators in \eqref{eq:detectordef} have spin two, the resulting tensor product splits into a scalar representation giving rise to the GEO discussed in~\eqref{eq:detector}, a spin-four representation and a mixed representation. Similar constructions can be performed for any spin $s$, but in higher dimensions the representation theory can become intricate.

In order not to clutter our formulas, we will restrict ourselves to the case of a four-dimensional bulk space ($d=2$) from now on. We can infer the correct normalization by going back to the definition of the operators $\phi^{(s)}_{J}(u_1,x^i)$ in terms of plane-wave creation operators of the free field expansion given in equation \eqref{eq:boundarybulk}. Indeed, choosing 
\begin{equation}
  \label{eq:13}
  \mathcal{C}_\Delta=\frac{2\Gamma(\Delta)e^{-\frac{\pi i\Delta}{2}}}{\pi K^2_{(s)}},
\end{equation}
we obtain the energy detector in its standard form 
\begin{equation}
  \label{eq:DGLAP}
 \Ed{0}{\Delta}(x)=\sum_{J}\int^{\infty}_0\frac{\dd \omega}{(2\pi)^{3}}\, \omega^{\Delta-1}\,(a^{(s)}_{J}(\omega,x))^\dagger a^{(s)}_{J}(\omega,x),
\end{equation}
where $a^{(s)}_J(\omega,x^i)$ annihilates a plane wave with momentum $\omega\, q(x^i)$.
Acting on a momentum eigenstate and taking into account the fundamental commutation relations \eqref{eq:commrel} this measures $(\textrm{energy})^{\Delta-2}$ deposited in a final state of helicity $J$
\begin{equation}
    \label{eq:actiononout}
    \Ed{0}{\Delta}(x) \ket{\omega q(x_1)}=\omega^{\Delta-2}\delta^{(2)}(x-x_1)\ket{\omega q(x_1)}.
\end{equation}

For $d=2$, detectors transform in representations of $\mathfrak{so}(2)$, which significantly simplifies the representation theory. In particular, we have the decomposition
\begin{equation}
 \mathbf{s}\otimes \mathbf{s}= \mathbf{0} \oplus \mathbf{0^-} \oplus \mathbf{2s} 
\end{equation}
 for any representations of spin $s$ of $\textrm{SO}(2)$, which allows us to treat particles of spin $s=1,2,\dots$ simultaneously. It is also useful to introduce a slightly different notation for these operators.

The generalized energy detector \eqref{eq:DGLAP} for particles of spin $s=1,2,\dots$ reads then
\begin{equation}
\label{eq:4dDGLAP}
    \mathcal{E}_{\Delta}(x^i)=\Ehel{++}{\Delta}(x^i)+\Ehel{--}{\Delta}(x^i),
  \end{equation}
  where we introduced the polarized detectors
  \begin{equation}
    \label{eq:polenergy}
    \Ehel{\pm\pm}{\Delta}=\int^{\infty}_0\frac{\dd \omega}{(2\pi)^{3}}\, \omega^{\Delta-1}\,\,(a^{(s)}_{\pm})^\dagger a^{(s)}_\pm,
  \end{equation}
in the energy basis.\footnote{Note that we suppress the spin of the detected particle not to clutter our notation. Thus, we denote the energy detectors of gluons and gravitons both by $\mathcal{E}_\Delta$. We hope that the meaning will be clear from the context.} These are simply defined by dropping the sum over the different polarizations in \eqref{eq:detector}.
Similarly, we can define the spinning detectors of helicity $\pm 2s$
\begin{align}
  \label{eq:hsenergy}
  \Ehel{+-}{\Delta}=\int^{\infty}_0\frac{\dd \omega}{(2\pi)^{3}}\, \omega^{\Delta-1}\,(a^{(s)}_+)^\dagger a^{(s)}_-,\qquad \Ehel{-+}{\Delta}=\int^{\infty}_0\frac{\dd \omega}{(2\pi)^{3}}\, \omega^{\Delta-1}\,(a^{(s)}_-)^\dagger a^{(s)}_+.
\end{align}
The operators $\Ehel{\pm\mp}{\Delta}(x^i)$ annihilate a particle of helicity $\mp s$ and create a particle of helicity $\pm s$, so that they have intrinsic helicity $\pm 2s$. The last operator is the pseudo-scalar representation
\begin{equation}
  \label{eq:9}
 \Ehel{0^-}{\Delta}(x^i)=\Ehel{++}{\Delta}(x^i)-\Ehel{--}{\Delta}(x^i),
\end{equation}
which transforms with a sign flip under the exchange $\pm\leftrightarrow\mp$.
As before, we suppressed a possible sum over an internal symmetry index.
\subsection{Detectors in the celestial basis}
In the above, we formulated energy operators in the plane wave basis \eqref{eq:DGLAP} and the position or Carrollian basis \eqref{eq:detector}. For completeness, we will also state their definition in terms of operators in the third basis \cite{Donnay:2022sdg,Donnay:2022wvx} that found application in recent years: the celestial basis \cite{Pasterski:2017kqt}. This appears to be a particularly natural choice since, like the detectors considered here, scattering states in that basis are labeled by a (boost) weight $\Delta$ and a representation of $\mathfrak{so}(2)$. More precisely, we relate the outgoing annihilation operator $a^{(s)}_{J}(\omega)$ to the celestial operator as
\begin{equation}
  \label{eq:celestial}
\mathcal{O}^{(s)}_J(\Delta,x^i)=\frac{1}{(2\pi)^{\frac{3}{2}}}\int^{\infty}_0\dd \omega\, \omega^{\Delta-1}\,a^{(s)}_{J}(\omega,x^i).
\end{equation}
While this choice of normalization appears to be non-standard, we find that it is consistent with the results below.

 We also define a conjugate celestial operator\footnote{Note that the underlying inner product in this case is the usual Fock-space norm for single particle states induced from the appropriate Klein--Gordon norm. This is distinct from other inner products that have been proposed in the context of celestial CFTs and that lead to different notions of conjugate operators; see, e.g., \cite{Crawley:2021ivb}.}
\begin{equation}
  \label{eq:5}
\bar{\mathcal{O}}^{(s)}_J(\Delta,x^i)=\frac{1}{(2\pi)^{\frac{3}{2}}}\int^{\infty}_0\dd \omega\, \omega^{\Delta-1}\,(a^{(s)}_{J}(\omega,x^i))^\dagger.
\end{equation}
Using the convolution formula for Mellin transforms,
\begin{equation}
  \label{eq:2}
  \int^{\infty}_0\dd \omega\, \omega^{\Delta-1}f(\omega) g(\omega)=\frac{1}{2\pi i}\int^{c+i\infty}_{c-i \infty}d s  f(s) g(\Delta-s),
\end{equation}
the analogue of \eqref{eq:detectordef} reads simply
\begin{equation}
  \label{eq:7}
\Ed{s}{\Delta}(x^i)=\frac{1}{2\pi i }\int^{c+i\infty}_{c-i \infty}d s\, \mathbb{P}^{(s)}[\bar{\mathcal{O}}^{(s_1)}(s,x^i)\mathcal{O}^{(s_1)}(\Delta-s,x^i)],
\end{equation}
where we again project onto an irreducible component inside the product $s_1\otimes s_1$.

In the next section, we will compute OPEs for GEOs \eqref{eq:DGLAP} and show that the spinning operators defined in \eqref{eq:hsenergy} naturally appear.

\section{Detector OPE from collinear limit}
\label{sec:OPE}
In this section, we aim to determine the leading term in the operator product expansion of the generalized energy operators for Einstein gravity and gauge theory. From now on, we will work exclusively in four bulk dimensions $(d=2)$. We begin with a brief review of how to obtain GEO correlators from Feynman diagram computations; more details can be found in~\cite{Caron-Huot:2022eqs,Herrmann:2024yai}.

\subsection{Energy correlators and weighted cross-sections}
Energy operators, or more generally detector operators, measure properties of the outgoing radiation at future null infinity for a given initial state. In other words, the quantity of interest is the expectation value of $n$ GEOs inserted on the celestial sphere,
\begin{equation}
  \label{eq:10}
  \mathbb{E}_n\sim \bra{\Psi}\mathcal{E}_{\Delta_1}\ldots \mathcal{E}_{\Delta_n}\ket{\Psi}.
\end{equation}
There are two natural options for preparing the initial state $\ket{\Psi}$. The first uses a local, gauge-invariant operator to create a state from the vacuum. The expectation value \eqref{eq:10} is then related to the \emph{form factors} of this operator. This choice is appropriate for CFTs \cite{Hofman:2008ar}. Another option is to take $\ket{\Psi}$ as an initial scattering state prepared at past null or timelike infinity. In the following, we adopt this latter perspective in which \eqref{eq:10} reads
\begin{equation}
  \label{eq:11}
  \mathbb{E}_n(\Delta_1,x_1;\ldots \Delta_n,x_n)=\frac{1}{N} \tensor[_{\rm in}]{\bra{p_1,\ldots p_m}}{}\mathcal{E}_{\Delta_1}(x_1)\ldots \mathcal{E}_{\Delta_n}(x_n)\tensor{\ket{p_1,\ldots p_m}}{_{\rm in}}.
\end{equation}
Here, we have suppressed any additional labels, such as spin or color, of the incoming particles. The normalization factor $N_m$ is necessary to deal with the formally appearing $\delta^{(4)}(0)$, which is familiar from standard cross-section computations. 

We will employ~\eqref{eq:11} to compute the leading terms in an OPE of GEOs. We will find that, in the limit $x_1\rightarrow x_2$, the $n$-point energy correlator $\mathbb{E}_n$ can be written as a coefficient times a lower-point correlator $\mathbb{E}_{n-1}$ with shifted weights. This allows us to extract the OPE between the operators $\mathcal{E}_{\Delta_1}(x_1)$ and $\mathcal{E}_{\Delta_2}(x_2)$. 

To perform this computation, it is convenient to reduce the object \eqref{eq:11} to a standard Feynman-diagram expansion as we sketch below; see \cite{Caron-Huot:2022eqs,Herrmann:2024yai} for further details.

The generalized energy operators discussed in the previous section are fully specified by their action on out-states~\eqref{eq:actiononout}.
The expectation value~\eqref{eq:11} is then simply computed by inserting a complete set of out-states 
\begin{equation}
    \label{eq:outunity}
   1=\sum^\infty_{k=0}\frac{1}{k!}\prod^{k}_{i=1}\int \frac{d^{3}\vec{k}_i}{(2\pi)^32k^0_i} \ket{k_1,\ldots,k_k}\bra{k_1,\ldots,k_k}\equiv \SumInt_{k}\ket{k_1,\ldots,k_k}\bra{k_1,\ldots,k_k}.
\end{equation}
For brevity, we suppress an additional sum over helicities and possible color labels. We also refrain from putting an explicit out-label on out-states.
Since these states are defined on future null infinity, we can simply evaluate the detector operators on them using~\eqref{eq:actiononout}. In order to avoid potential contact terms, we assume that all energy detectors are inserted at distinct points on the celestial sphere. Taking into account the proper, and assuming that all particles are indistinguishable and bosonic, one finds then 
\begin{multline}
  \label{eq:final}
  \mathbb{E}_n=\frac{1}{N_m} \prod^{n}_{a=1}\int^{\infty}_0\frac{\dd \omega_a}{(2\pi)^3}\omega^{\Delta_a-1}_a\SumInt_k(2\pi)^4\delta^{(4)}(\omega_1 q(x_1)+\ldots +\omega_n q(x_n)+K-P)|\mathcal{M}_{m\rightarrow k+n}|^2\,,
\end{multline}
where $K=\sum^{k}_{i=1} k_i, P=\sum^m_{i=1} p_i$, the matrix element is
\begin{equation}
  \label{eq:16}
  \tensor{\braket{k_1,\ldots k_k,\omega_1q(x_1),\ldots \omega_nq(x_n)|p_1,\ldots p_m}}{_{\rm in}}=(2\pi)^4\delta^{(4)}(\omega_1 q(x_1)+\ldots \omega_n q(x_n)+K-P)\mathcal{M}_{m\rightarrow k+n},
\end{equation}
and the remaining normalization factor depends only on the incoming states $p_i$. 
In the squared matrix element $|\mathcal{M}_{m\rightarrow k+n}|^2$ we sum over the polarizations of all detected and undetected particles
\begin{equation}
\label{eq:polsum}
 |\mathcal{M}_{m\rightarrow k+n}|^2=(\mathcal{M}_{m\to k+n})^* \Pi(q_1)\ldots \Pi(q_n) \Pi(k_1)\ldots \Pi(k_k)   \mathcal{M}_{m\rightarrow k+n},
\end{equation}
where $\Pi(q)$ denotes the projector on the physical degrees of freedom, and we suppressed the spacetime indices for clarity.
Color sums over the particles are to be included as well if necessary.

We note in passing that one can write the above in a more covariant form by introducing a detector vertex factor and delta functions that put every undetected particle on-shell. In this way, the correlator can be constructed in a standard Feynman-rule fashion in the in-in formalism to any order in perturbation theory. We refer to \cite{Caron-Huot:2022eqs, Chang:2025zib} for more examples in the context of scalar theories and Yang--Mills, and to \cite{Herrmann:2024yai} for the first computation of the leading terms of a two-point correlator in Einstein gravity and an elucidating diagrammatic expansion of \eqref{eq:final}.

Equation~\eqref{eq:final} can be easily modified to compute correlation functions of the other detectors introduced in \eqref{eq:polenergy} and \eqref{eq:hsenergy}. 
The GEO \eqref{eq:4dDGLAP} measures the energy deposited in both polarizations of a gluon or graviton in the final state which leads to the appearance of the projectors $\Pi_{\mu\nu}$ or $\Pi_{\mu\nu\alpha\beta}$ in~\eqref{eq:polsum}. To compute correlators of the other operators, one simply needs to take into account the different polarization structures.
Defining the correlator
\begin{equation}
  \label{eq:19}
  \mathbb{E}^{\rho_1\sigma_1\ldots \rho_n\sigma_n}_n(\Delta_1,x_1;\ldots \Delta_n,x_n)=\frac{1}{N} \tensor[_{\rm in}]{\bra{p_1,\ldots p_m}}{}\mathcal{E}^{\rho_1\sigma_1}_{\Delta_1}(x_1)\ldots \mathcal{E}^{\rho_n\sigma_n}_{\Delta_n}(x_n)\tensor{\ket{p_1,\ldots p_m}}{_{\rm in}},
\end{equation}
where $\rho_i,\sigma_i=\pm$
we find the general formula
\begin{equation}
\begin{split}
  \mathbb{E}^{\rho_1\sigma_1\ldots \rho_n\sigma_n}_n=\frac{1}{N_m} \prod^{n}_{a=1}&\int^{\infty}_0\frac{\dd \omega_a}{(2\pi)^3}\omega^{\Delta_a-1}_a\SumInt_k(2\pi)^4\delta^{(4)}(\omega_1 q(x_1)+\ldots \omega_n q(x_n)+K-P)\\
                                    &\qquad \mathcal{M}^{\mu_1\nu_1\ldots\mu_n\nu_n}_{m\rightarrow k+n}(\epsilon^{\sigma_1}_{\mu_1\nu_1})^*\ldots (\epsilon^{\sigma_n}_{\mu_n\nu_n})^*\epsilon^{\rho_1}_{\alpha_1\beta_1}\ldots\epsilon^{\rho_n}_{\alpha_n\beta_n}(\mathcal{M}^{\alpha_1\beta_1\ldots\alpha_n\beta_n}_{m\rightarrow k+n})^*,
                                      \label{eq:finalpol}
\end{split}
\end{equation}
where the polarization tensors $\epsilon^{\sigma_k}_{\mu_k\nu_k}$ are associated to the null momentum $q^\mu(x_k)$ and we suppressed an additional sum over the polarization of the undetected particles. The case for gauge theory proceeds in complete analogy.
\subsection{Detector OPE from collinear factorization}
We are now in the position to study the OPE of GEOs for Einstein gravity using \eqref{eq:final}. In the limit $x_1\rightarrow x_2$, the two null vectors $q^\mu(x_1)$ and $q^\mu(x_2)$ become collinear. In this limit, a gravitational scattering amplitude factorizes as 
\begin{align}
  \label{eq:18}
  \mathcal{M}_n(\ldots;\omega_1q_1,\sigma_1;\omega_2q_2,\sigma_2)=\sum_\sigma{\rm Split}_{-\sigma}(\sigma_1,\sigma_2)\mathcal{M}^{\mu\nu}_{n-1}(\ldots; K,\sigma)(\epsilon^{\sigma}_{\mu\nu})^*+\ldots
\end{align}
where $K^\mu=\omega_1q^\mu_1+\omega_2q^\mu_2$ and the splitting function is defined in terms of the gravitational three-point function $V$ as
\begin{equation}
  \label{eq:36}
 {\rm Split}_{-\sigma}(\sigma_1,\sigma_2)=\frac{1}{K^2}(\epsilon^{\mu_1\nu_1}_{\sigma_1})^*(\epsilon^{\mu_2\nu_2}_{\sigma_2})^*V_{\mu_1\nu_1\mu_2\nu_2\mu\nu}(\omega_1q_1,\omega_2q_2,-K)\epsilon^{\mu\nu}_\sigma,
\end{equation}
Equation \eqref{eq:polsum} becomes then in the limit
\begin{align}
  \label{eq:splitting}
&\sum_{\sigma_1,\sigma_2}|\mathcal{M}_{m\rightarrow k+n}|^2\rightarrow \frac{1}{2}\sum_{\sigma_1,\sigma_2,\sigma=\pm}|{\rm Split}_{-\sigma}(\sigma_1,\sigma_2)|^2 \sum_{\sigma'}|\mathcal{M}_{m\rightarrow k+n-1}(\sigma',K)|^2 +\\
  &\qquad\sum_{\sigma_1,\sigma_2,\sigma=\pm}{\rm Split}_{\sigma}(\sigma_1,\sigma_2) \overline{{\rm Split}}_{-\sigma}(\sigma_1,\sigma_2)\mathcal{M}_{m\rightarrow k+n-1}(\ldots; K,-\sigma)(\mathcal{M}_{m\rightarrow k+n-1}(\ldots; K,\sigma))^*,\nonumber
\end{align}
where again we have suppressed an additional sum over the undetected particles.
This equation can be regarded as decomposing the symmetric product of two spin two representations into a scalar representation in the first line and, in the last line, the symmetric traceless representations of helicity $\pm 4$.
According to the discussion in Section \ref{sec:detectorsdef}, we can expect that the first line gives rise to the usual GEO \eqref{eq:4dDGLAP}, while the second line gives rise to contributions coming from spinning detectors \eqref{eq:hsenergy}.

Plugging this back into \eqref{eq:final}, we obtain two contributions
\begin{equation}
  \label{eq:matching}
\lim_{x_1\rightarrow x_2}\mathbb{E}_{n-1}\equiv \mathbb{E}_{\rm diag}+\mathbb{E}_{\rm off-diag}\,.
\end{equation}
The diagonal contribution from the first line of \eqref{eq:splitting} yields
\begin{equation}
\begin{split}
  \mathbb{E}_{\rm diag}=\frac{1}{2} \int^{\infty}_0 d\omega_1 d\omega_2 \, \omega_2^{\Delta_2-1}  \, &\omega_1^{\Delta_1-1} \sum_{\sigma_1,\sigma_2,\sigma=\pm}|{\rm Split}_{\sigma}(\sigma_1,\sigma_2)|^2\frac{1}{N_m} \prod^{n}_{a=3}\int^{\infty}_0\frac{\dd \omega_a}{(2\pi)^3}\omega^{\Delta_a-1}_a \\
  &\SumInt_k(2\pi)^4\delta^{(4)}(\omega q(x_2)+\ldots \omega_n q(x_n)+K-P)|\mathcal{M}_{m\rightarrow k+n-1}|^2\,,
\end{split}
\end{equation}
With the change of variables $\omega_1=t\omega, \omega_2=(1-t)\omega$ we obtain
\begin{equation}
  \label{eq:23}
  \mathbb{E}_{\rm diag}= \mathscr{C} \, \mathbb{E}^{(0)}_{n-1}(\Delta_1+\Delta_2,x_2;\cdots;\Delta_{n},x_n),
\end{equation}
where the coefficient is
\begin{equation}
  \label{eq:28}
  \mathscr{C}(\Delta_1,\Delta_2,z_{12},\bar{z}_{12})=\frac{1}{2}\int^{1}_0\, dt \,t^{\Delta_2-1}  \, (1-t)^{\Delta_1-1} \sum_{\sigma_1,\sigma_2,\sigma=\pm}|{\rm Split}_{\sigma}(\sigma_1,\sigma_2)|^2,
\end{equation}
and we used the fact that the splitting functions in gravity are independent of energy. Note that $t$ has the interpretation of the fraction of energy transmitted to particles $1$ and $2$. We will use this observation in the next section. 

Similarly, from comparing with \eqref{eq:finalpol}, we see that the off-diagonal term gives rise to a correlation function of detectors with spin
\begin{equation}
  \label{eq:29}
  \mathbb{E}_{\rm off-diag}=\sum_{\sigma,\sigma'}\mathscr{C}_{\sigma,\sigma'}\mathbb{E}^{\sigma\sigma'}(\Delta_1+\Delta_2,\ldots, \Delta_{n}),
\end{equation}
with coefficient
\begin{equation}
  \label{eq:30}
  \mathscr{C}_{\pm\mp}(\Delta_1,\Delta_2,z_{12},\bar{z}_{12})= 2\int^1_0 dt \,t^{\Delta_1-1}  \, (1-t)^{\Delta_2-1} {\rm Split}_{\pm}(-,+) {\rm Split}_{\pm }(+,-).
\end{equation}
The gravitational splitting function in our convention has the form
\begin{equation}
  \label{eq:31}
  {\rm Split}_{+}(-,-)=\frac{\kappa}{t(1-t)}\frac{\zbar_{12}}{z_{12}},\qquad  {\rm Split}_{+}(+,-)=\frac{\kappa (1-t)^3}{t}\frac{z_{12}}{\zbar_{12}},\qquad  {\rm Split}_{+}(-,+)=\frac{\kappa t^3}{1-t}\frac{z_{12}}{\zbar_{12}},
\end{equation}
with $\kappa^2=32\pi G$.
From~\eqref{eq:matching} we can read off the leading term of the gravitational detector OPE
\begin{align}
  \label{eq:gravitydetector}
\mathcal{E}_{\Delta_1}(x_1)\mathcal{E}_{\Delta_2}(x_2)\overset{x_1\rightarrow x_2}{\sim}c_0 \mathcal{E}_{\Delta_1+\Delta_2}(x_2)+c_+ \frac{\zbar^2_{12}}{z^2_{12}}\Ehel{-+}{\Delta_1+\Delta_2}(x_2)+c_- \frac{z^2_{12}}{\zbar^2_{12}}\Ehel{+-}{\Delta_1+\Delta_2}(x_2)+\ldots
\end{align}
where the coefficients are given by
\begin{subequations}
\begin{align}
  \label{eq:33}
  c_0&=\kappa^2  [B(\Delta_1-2,\Delta_2-2)+B(\Delta_1-2,\Delta_2+6)+B(\Delta_1+6,\Delta_2-2)]\\
  c_{\pm}&=2\kappa^2 B(\Delta_1+2,\Delta_2+2),
\end{align}
\end{subequations}
and the operators $\Ehel{\pm\mp}{\Delta}$ are the spinning energy detectors of helicity $\pm 4$ introduced in~\eqref{eq:hsenergy}. We therefore see that the OPE naturally leads one to consider GEOs and their spinning versions even if one starts out with the usual energy operator only.

In the same way, one can also compute the leading OPE terms between the other operators defined at the end of Section~\ref{sec:detectorsdef}. We find for the spinning operators with helicity four
\begin{subequations}
\begin{align}
  \label{eq:34}
  \Ehel{+-}{\Delta_1}(x^i_1)\Ehel{+-}{\Delta_2}(x^i_2)&\sim \kappa^2B(\Delta_1-2,\Delta_2-2)\frac{\zbar^2_{12}}{z_{12}^{ 2}}
  \Ehel{+-}{\Delta_1+\Delta_2}(x^i_2), \\
  \Ehel{+-}{\Delta_1}(x^i_1)\Ehel{-+}{\Delta_2}(x^i_2)&\sim \kappa^2\left( B(\Delta_1+2,\Delta_2+2)\mathcal{E}_{\Delta_1+\Delta_2}+B(\Delta_1+6,\Delta_2-2)\frac{z^2_{12}}{\zbar^2_{12}}\Ehel{+-}{\Delta_1+\Delta_2}\right.\\
&\qquad\qquad\qquad\qquad\qquad\qquad\qquad\qquad\left.+B(\Delta_1-2,\Delta_2+6)\frac{\zbar^2_{12}}{z^2_{12}}\Ehel{-+}{\Delta_1+\Delta_2}\right)\nonumber
\end{align}
\end{subequations}
and for the polarized detectors
\begin{subequations}
\begin{align}
  \label{eq:35}
  \Ehel{++}{\Delta_1}(x^i_1)\Ehel{++}{\Delta_2}(x^i_2)&\sim \kappa^2B(\Delta_1-2,\Delta_2-2)\Ehel{++}{\Delta_1+\Delta_2}(x^i_2),\\
  \Ehel{++}{\Delta_1}(x^i_1)\Ehel{--}{\Delta_2}(x^i_2)&\sim \kappa^2\left(B(\Delta_1-2,\Delta_2+6)\Ehel{++}{\Delta_1+\Delta_2}+B(\Delta_1+6,\Delta_2-2)\Ehel{--}{\Delta_1+\Delta_2}\right.\\
 &\qquad\qquad\left.+B(\Delta_1+2,\Delta_2+2)\frac{z^2_{12}}{\zbar^2_{12}}\Ehel{+-}{\Delta_1+\Delta_2}+B(\Delta_1+2,\Delta_2+2)\frac{\zbar^2_{12}}{z^2_{12}}\Ehel{-+}{\Delta_1+\Delta_2}\right)\nonumber
\end{align}
\end{subequations}
We therefore see that the class of operators discussed in Section \ref{sec:detectorsdef} is closed to leading order in the OPE.

We have framed the above discussion in terms of gravitational detectors. But since the leading order is completely fixed by the splitting functions through equation \eqref{eq:splitting}, it is straightforward to obtain the leading OPEs for detector operators in other theories. In particular, we can reproduce the case of Yang--Mills that was worked out already in \cite{Chen:2021gdk}. We find
\begin{align}
  \label{eq:YMdetector}
\mathcal{E}_{\Delta_1}(x_1)\mathcal{E}_{\Delta_2}(x_2)\overset{x_1\rightarrow x_2}{\sim}\frac{c_0}{|z_{12}|^2} \mathcal{E}_{\Delta_1+\Delta_2-2}(x_2)+ \frac{c_+}{z^2_{12}}\Ehel{-+}{\Delta_1+\Delta_2-2}(x_2)+ \frac{c_-}{\zbar^2_{12}}\Ehel{+-}{\Delta_1+\Delta_2-2}(x_2)+\ldots
\end{align}
where the coefficients are given by
\begin{subequations}
\begin{align}
  \label{eq:33b}
  c_0&=2g^2_{\rm YM}C_A  [B(\Delta_1-2,\Delta_2)+B(\Delta_1,\Delta_2-2)+B(\Delta_1,\Delta_2)]\\
  c_{\pm}&=2g^2_{\rm YM}C_A B(\Delta_1,\Delta_2).
\end{align}
\end{subequations}
Here, the operators $\Ehel{\pm\mp}{\Delta}(x_2)$ are the spinning gluon detectors of helicity $\pm 2$, and $C_A$ is the Casimir in the adjoint representation. In contrast to the gravitational detectors \eqref{eq:gravitydetector}, the leading contribution to the OPE is divergent \cite{Chen:2021gdk}, which is related to the presence of collinear singularities in Yang--Mills theory. As in the gravity case, one can also straightforwardly compute the OPE of spinning operators using equation~\eqref{eq:finalpol}.

In this section, we determined the leading terms of OPEs of generalized energy detectors using the universal splitting functions. However, these results will receive corrections at subleading order both in the collinear expansion and in the coupling constant. For instance, the collinear behavior of two gravitational energy detectors in the specific case of a two-point function in the state prepared by two massless scalar fields coupled to gravity was worked out to one-loop order in \cite{Herrmann:2024yai}. The general form of these corrections should be constrained by the two-dimensional Euclidean conformal group.

We conclude this section by drawing attention to the case of self-dual gravity. This theory propagates only gravitons of, say, positive helicity, so that the only nontrivial detector is the polarized detector $\Ehel{++}{\Delta}$. The OPE of this detector was determined in \eqref{eq:35}. The amplitudes of this theory are known exactly and exhibit an infinite-dimensional $w_{1+\infty}$ symmetry \cite{Ball:2021tmb} that should fix the form of the subleading terms. This theory would therefore provide a very interesting toy model for studying the behavior of generalized energy correlators in more detail. We leave this to future work.

\section{Detector operators and celestial holography}
\label{sec:celestial}
In the previous section, we computed the OPE between GEOs to leading order in the coupling constant and in the collinear limit. In this section, we want to make contact with operator product expansions known from the context of celestial holography. 

Furthermore, conformally soft operators have been defined in the context of celestial holography, governing the low-energy behavior of massless gauge theories~\cite{Strominger:2013lka,Strominger:2013jfa,Donnay:2018neh,Pasterski:2021dqe}. We will argue that analogous \emph{soft} energy operators can be considered.
\subsection{Detector OPE from celestial OPE}
Celestial OPEs have been derived from various points of view: from the collinear limits of scattering amplitudes~\cite{Fan:2019emx,Fotopoulos:2019vac}; from conformally soft theorems \cite{Pate:2019lpp}; or just from the requirements of Poincaré invariance~\cite{Himwich:2021dau}.
The OPE between two outgoing graviton operators with positive helicity $O^{(2)}_+$ is given by
\begin{equation}
  \label{eq:celestialOPE}
O^{(2)}_+(\Delta_1,z_1,\zbar_1)O^{(2)}_+(\Delta_2,z_2,\zbar_2)\sim -\frac{\kappa}{2}\frac{\zbar_{12}}{z_{12}}B(\Delta_1-1,\Delta_2-1)O^{(2)}_+(\Delta_1+\Delta_2,z_2,\zbar_2)+\ldots,
\end{equation}
where it is assumed that one takes the holomorphic limit $z_{12}\rightarrow 0$, keeping $\bar{z}_{12}$ fixed. Since this OPE can be derived from the collinear limits of scattering amplitudes, it is expected to be compatible with the detector OPEs derived in the last section. Indeed, taking the appropriate contractions  and using the first Mellin--Barnes lemma, we find
\begin{align}
  \label{eq:22}
  \Ehel{++}{\Delta_1}(x^i_1)\Ehel{++}{\Delta_2}(x^i_2)&=\frac{1}{(2\pi i)^2}\int^{c_0+i\infty}_{c_0-i\infty} \!\!\!\!\!ds\, dt\, \wick{ \c1{O}^{(2)}_+(s,x^i_1)    \c2{\bar{O}}^{(2)}_+(\Delta_1-s,x^i_1)  \c1{O}^{(2)}_+(t,x^i_2)  \c2{\bar{ O}}^{(2)}_+(\Delta_2-t,x^i_2)}\nonumber\\
  &=\kappa^2B(\Delta_1-2,\Delta_2-2)\Ehel{++}{\Delta_1+\Delta_2}(x^i_2)
\end{align}
in agreement with equation \eqref{eq:35}.

Similarly, the OPE between two positive-helicity gluons is
\begin{equation}
  \label{eq:47}
  O^{(1)a}_+(\Delta_1,z_1,\zbar_1)O^{(1)b}_+(\Delta_2,z_2,\zbar_2)\sim\frac{-igf\indices{^{ab}_c}}{z_{12}}B(\Delta_1-1,\Delta_2-1)O^{(1)c}_+(\Delta_1+\Delta_2-1,z_2,\zbar_2).
\end{equation}
Using the same identity as above, we find the correlator for polarized energy detectors
\begin{equation}
  \label{eq:49}
\Ehel{++}{\Delta_1}(x^i_1)\Ehel{++}{\Delta_1}(x^i_2)=\frac{2g^2_{\rm YM}C_A}{|z_{12}|^2}B(\Delta_1-2,\Delta_2-2)\Ehel{++}{\Delta_1+\Delta_2-2}(x^i_2),
\end{equation}
compatible with the contribution of the polarized detector OPE to the full OPE, captured by the last term in \eqref{eq:33b}.

\subsection{The emergence of the number operator}
\label{sec:softops}
As noted before, conservation of energy in a three-point vertex can be kinematically interpreted as a one-to-two splitting process, each of them carrying energy $  t\omega$ and $(1-t)\omega$, respectively.
In these variables, the soft limit is captured by the integration region $t\sim 0$ (or, equivalently, $t\sim 1$). At the level of the OPE coefficient $ \mathscr{C}$ in \eqref{eq:28}, its leading singularity comes from 
\begin{equation}
\left. \mathscr{C} \right|_{t\to0} \propto\frac{1}{|z_{12}|^{2(2-s)}} \frac{1}{\Delta-2}+\cdots\,,
\end{equation}
where $s=2$ for gravitons and $s=1$ for gluons. The interesting point is that for any helicity $s$ the soft limit is dominated by the operator detector with conformal dimension $\Delta=2$. This case corresponds to a particle-counting detector. More precisely, we can define the operator
\begin{equation}
\label{eq:softdetect}
\mathcal{N}(z,\bar{z})=\lim_{\Delta\to 2} (\Delta-2)\mathcal{E}_\Delta(z,\bar{z})\,.
\end{equation}
Taking this limit directly in the detector OPEs \eqref{eq:gravitydetector} and \eqref{eq:YMdetector} we find
\begin{equation}
\label{OPE:N}
\mathcal{N}(z_1,\bar{z}_1)\mathcal{E}_{\Delta}(z_2,\bar{z}_2)\sim\frac{g_s^2}{|z_{12}|^{2(2-s)}} \mathcal{E}_{\Delta+2(s-1)}(z_2,\bar{z}_2)
\end{equation}
where $g_2=\kappa$ in Einstein gravity and $g_1^2=2 g^2_{\rm YM} C_A$ for pure Yang-Mills theory.

This operator will play an important role in the IR sector of celestial holography. In the next section, we will evaluate its expectation value. Notably, the application of the conservation of the asymptotic soft charge \cite{Strominger:2013jfa} will show that $\mathcal{N}$ is made out of a product of conformally soft currents.

Before examining $\mathcal{N}$ more closely, we note that the OPE coefficients for gauge theory and gravity have poles at any integer value $\Delta\le 2$. In analogy with the number operator \eqref{eq:softdetect} we can therefore formally define
\begin{equation}
    \mathcal{N}^{(n)}(x^i)=\lim_{\epsilon\to 0 }\epsilon\, \mathcal{E}_{2-n+\epsilon}(x^i).
\end{equation}
We comment on the interpretation of this particular set of operators in the concluding remarks in Section~\ref{sec:conclusion}.

\section{Detector correlators and soft properties}
\label{sec:softcor}
The OPE of GEOs, both in Yang--Mills theory \eqref{eq:YMdetector} and gravity \eqref{eq:gravitydetector}, has poles when either detector has weight $\Delta=2$. Naively, the corresponding operators $\mathcal{E}_{\Delta=2}$ weight every outgoing state by $\omega^0$ and thus can be viewed as counting the number of outgoing particles. In the context of QCD, this operator was carefully studied in \cite{Chang:2025zib}. Here, we want to show that this operator can also be understood using the asymptotic symmetries of gauge theories in asymptotically flat spacetimes. We will first consider the case of gravity before discussing the analogous situation in Yang--Mills theory.

\subsection{The \texorpdfstring{$\Delta=2$}{Delta=2} operator in gravity}
\label{sec:op2ingravity}
Consider the one-point function of a gravitational GEO in a generic scattering state. Using equation~\eqref{eq:final}, we can write this as
\begin{equation}
  \label{eq:21}
  \mathbb{E}(\Delta,x^i)=\frac{1}{N_m}\int^{\infty}_0\frac{\dd \omega}{(2\pi)^3}\omega^{\Delta-1}\SumInt_k  (2\pi)^4\delta^{(4)}(\omega q(x^i)+K-P)|\mathcal{M}_{m\rightarrow k+1}|^2.
\end{equation}
Projecting on the pole at $\Delta=2$ of this correlator we have
\begin{align}
  \label{eq:32}
  \lim_{\Delta\rightarrow 2}(\Delta-2) \mathbb{E}(\Delta,x^i)=\frac{1}{N_m}\int^{\infty}_{-\infty}\frac{\dd \omega}{(2\pi)^3}\delta(\omega)\SumInt_k  (2\pi)^4\delta^{(4)}(\omega q(x^i)+K-P)\omega^2|\mathcal{M}_{m\rightarrow k+1}|^2,
\end{align}
where we used the representation
\begin{equation}
  \label{eq:37}
  \lim_{\epsilon\rightarrow 0}\epsilon |x|^{\epsilon-1}=2\delta(x).
\end{equation}
This appearance of the delta function shows that we focus on the sector where the graviton captured by the detector operator is soft and we can project onto the soft pole of the scattering amplitude.
The soft theorem for an outgoing graviton of polarization tensor $\epsilon_{\mu\nu}$ reads
\begin{equation}
  \label{eq:softtheorem}
 \lim _{\omega\to 0}\mathcal{M}_{m\rightarrow n+1}=\frac{\kappa}{2}\left[\sum_{k\in {\rm out}}\frac{\epsilon_{\mu\nu} p^\mu_{k}p^\nu_k}{p_{k}\cdot q}-\sum_{k\in {\rm in}}\frac{\epsilon_{\mu\nu} p^\mu_{k}p^\nu_k}{p_{k}\cdot q}\right]\mathcal{M}_{m+n}.
\end{equation}
Squaring and summing over the polarizations of the outgoing gravitons since the detector is sensitive only to the sum, we find
\begin{equation}
    \lim_{\omega \to 0}\omega^2 |\mathcal{M}_{m\rightarrow n+1}|^2=\frac{\kappa^2}{4}\left[\sum_{i,j\in {\rm in}}-2 \sum_{i\in {\rm out},j\in {\rm in}}+\sum_{i,j\in {\rm out}}\right]\omega_i\omega_j K^{(2)}(x_i,x_j,x)|\mathcal{M}_{m+n}|^2,
\end{equation}
where
\begin{equation}
\label{eq:eikonalkernel}
    K^{(s)}(x_i,x_j,x)=\frac{(q_i\cdot q_j)^{s}}{q_i\cdot q\, q_j \cdot q}=\frac{|z_1-z_2|^{2s}}{|z-z_1|^2 |z-z_2|^2}, \qquad q_{i,j}=q(x_{i,j}), q=q(x),
\end{equation}
is the eikonal kernel,
and we used the symmetry of the kernel to simplify the result. We therefore obtain
\begin{align}
    \lim_{\Delta\rightarrow 2}(\Delta-2)\mathbb{E}_{\Delta}(x)=\frac{\kappa^2(2\pi)^4}{4N_m}\SumInt_{k}\delta^{(4)}(K-P) \left[\sum_{i,j\in {\rm in}}-2\!\!\! \sum_{i\in {\rm out},j\in {\rm in}}\!\!\!+\sum_{i,j\in {\rm out}}\right]\omega_i\omega_j  K(x_i,x_j,x)|\mathcal{M}_{m+n}|^2.
\end{align}
Note that we can equivalently write this as
\begin{align}
    \lim_{\Delta\rightarrow 2}(\Delta-2)\mathbb{E}_{\Delta}(x)&=\frac{\kappa^2}{4N_m}\int d^2 y_1 d^2y_2  K(y_1,y_2,x) \\
    &\bra{p_1\ldots p_m}\left[(\mathcal{E}(y_1)-\mathcal{E}^{\rm in}(y_1))(\mathcal{E}(y_2)-\mathcal{E}^{\rm in}(y_2))\right]\ket{p_1\ldots p_m}\nonumber,
\end{align}
where $\mathcal{E}^{\rm in}(x)$ denotes energy detectors inserted at past null infinity acting directly on the in-states.

We can now make direct contact with the extensive literature on asymptotic symmetries. On future null infinity, the charge for supertranslations $Q[\mathcal{T}]$ is
\begin{equation}
  \label{eq:39}
    Q[\mathcal{T}]=\int d^2x \mathcal{T}\left(\tfrac{1}{2}\mathcal{E}(x)-q_{\rm soft}(x)\right),
  \end{equation}
  parametrized by $\mathcal{T}(x^i)$. It is given by the outgoing (hard) energy flux measured by $\mathcal{E}(x)\equiv \mathcal{E}_{\Delta=3}(x)$ together with a soft piece defined as \cite{Strominger:2013jfa,He:2014laa}
  \begin{equation}
    \label{eq:40}
    q_{\rm soft}(x)=\frac{2}{\kappa^2}\int du\, (\partial^2_z \partial_uC_{\bar{z}\bar{z}}+\partial^2_{\bar{z}} \partial_uC_{zz}).
  \end{equation}
  Here, $C_{zz}$ is the shear that carries the two polarizations of the gravitational field.\footnote{We neglect an additional term in the charge that ensures the correct transformation under superrotation if one considers the extended BMS group \cite{Barnich:2011mi,Donnay:2021wrk,Campiglia:2021bap}.} In the notation of Section \ref{sec:detectorsdef}, it corresponds to the Carrollian field with spin $s=2$. A similar expression for the charge can be obtained in terms of quantities defined on past null infinity. Conservation of the supertranslation charge then states that these two expressions are equal. In other words, the energy flux is conserved at every angle due to soft contributions. In terms of the integrands, we have therefore
\begin{equation}
    \label{eq:chargeconsaspect}
    \tfrac{1}{2}(\mathcal{E}^{\rm out}(x)-\mathcal{E}^{\rm in}(x))=q_{\rm soft}(x)-q^{\rm in}_{\rm soft}(x).
  \end{equation}
  It is convenient to rewrite the right-hand side in terms of the supertranslation current introduced in \cite{Strominger:2013jfa} as
\begin{equation}
    \label{eq:chargeconsaspectb}
    \mathcal{E}^{\rm out}(x)-\mathcal{E}^{\rm in}(x)=\frac{\partial_{\bar{z}} P_z}{\pi}.
  \end{equation}
   We have therefore the compact relation
  \begin{equation}
    \label{eq:41}
    \lim_{\Delta\rightarrow 2}(\Delta-2)\mathbb{E}_{\Delta}(x)=\frac{\kappa^2}{4\pi^2N_m}\int d^2 y_1 d^2y_2  K^{(2)}(y_1,y_2,x)\bra{p_1\ldots p_m} \partial_{\bar{z}}P_z(y_1)\partial_zP_{\zbar}(y_2)\ket{p_1\ldots p_m},
  \end{equation}
  where we used $\partial_{\bar{z}}P_z=\partial_zP_{\zbar}$ due to the boundary conditions obeyed by the gravitational field close to spatial infinity \cite{Strominger:2013jfa}. Note that this identity holds to all orders in perturbation theory since the leading soft-graviton theorem~\eqref{eq:chargeconsaspectb} does not receive loop-corrections~\cite{Bern:2014oka}.
  
  We now make direct contact with the discussion in Section \ref{sec:softops}. The soft current $P_z$ is related to the celestial graviton operator by \cite{Pate:2019lpp}
  \begin{equation}
    \label{eq:42}
P_z=-\frac{2}{\kappa}\lim_{\Delta\to 1}(\Delta-1)\partial_{\bar{z}}O^{(2)}_+(\Delta).
  \end{equation}
With this relation and the distributional identity
\begin{equation}
  \label{eq:44}
\partial^2_{\zbar_1}\partial^2_{z_2}K^{(2)}(x_1,x_2,x)=4 \frac{(z_1-z)(\zbar_2-\zbar)}{(\zbar_1-\zbar)^3(z_2-z)^3}+\pi^2 \delta^{(2)}(x-x_1)\delta^{(2)}(x-x_2),
\end{equation}
we find for \eqref{eq:41}
\begin{equation}
\label{eq:almosttheregravity}
\begin{split}
  &\lim_{\Delta\rightarrow 2}(\Delta-2)\mathbb{E}_{\Delta}(x)=\lim_{\Delta_1,\Delta_2\to 1}(\Delta_1-1)(\Delta_2-1)\Big[\frac{1}{N_m}\bra{p_1\ldots p_m}O^{(2)}_{+}(\Delta_1,x)O^{(2)}_{-}(\Delta_2,x)\ket{p_1\ldots p_m}\\
  &\qquad\qquad +\frac{4}{\pi^2N_m}\int d^2x_1d^2 x_2 \frac{(z_1-z)(\zbar_2-\zbar)}{(\zbar_1-\zbar)^3(z_2-z)^3}\bra{p_1\ldots p_m}O^{(2)}_{+}(\Delta_1,x_1)O^{(2)}_{-}(\Delta_2,x_2)\ket{p_1\ldots p_m}\Big].
  \end{split}
\end{equation}
We recognize the terms in the second line as shadow transforms of the respective operators where we use the normalization and conventions of \cite{Osborn:2012vt}. Note that the shadow transform of $O^{(2)}_{\pm}(\Delta_1=1,x)$ has the same quantum numbers as $O^{(2)}_{\mp}(\Delta_1=1,x)$. As shown in detail in \cite{Pasterski:2021fjn,Pasterski:2021dqe}, conformally soft operators are organized in a special multiplet of the Lorentz group such that these operators are indeed just shadow transforms of one another. In other words, the shadow transforms the soft factor for an outgoing positive helicity graviton in \eqref{eq:softtheorem} into the one for a negative helicity graviton.
The second term in~\eqref{eq:almosttheregravity} is therefore proportional to the first. Taking the normalization into account, we arrive at our final result:
\begin{equation}
\begin{split}
\label{eq:finalgrav}
\lim_{\Delta\rightarrow 2}(\Delta-2)\mathbb{E}_{\Delta}(x)=&\lim_{\Delta_1,\Delta_2\to 1}(\Delta_1-1)(\Delta_2-1)\\
&\frac{1}{N_m}\bra{p_1\ldots p_m}O^{(2)}_{+}(\Delta_1,x)O^{(2)}_{-}(\Delta_2,x)+O^{(2)}_{-}(\Delta_1,x)O^{(2)}_{+}(\Delta_2,x)\ket{p_1\ldots p_m}.
\end{split}
\end{equation} 
This allows us to read off the operator identity
\begin{equation}
   \mathcal{N}=\lim_{\Delta_1,\Delta_2\to 1}(\Delta_1-1)(\Delta_2-1)\left(O^{(2)}_{+}(\Delta_1)O^{(2)}_{-}(\Delta_2)+O^{(2)}_{-}(\Delta_1)O^{(2)}_{+}(\Delta_2)\right).
\end{equation}

Let us note that this representation of the number operator allows for independent consistency checks.\footnote{Formally, this result can also be derived from the celestial representation~\eqref{eq:7} of the GEO by contour integration. However, in the absence of a complete understanding of the pole structure of celestial operators, this argument remains heuristic.} For instance, in the self-dual case, one can compute independently the OPE of the number operator and the detector itself. In fact, using the soft theorem in the positive (negative) helicity sector,
we holographically recover the OPE of the number operator \eqref{OPE:N}. A completely analogous treatment applies to self-dual YM, as we will see below.

\subsection{The \texorpdfstring{$\Delta=2$}{Delta=2} operator in gauge theory}
We now turn to the analogous situation in Yang--Mills theory. The $\Delta=2$-pole of the GEO has been studied in great detail in \cite{Chang:2025zib}. In contrast to the former work that analyzed the one-point function in the background of a state created by a local, gauge-invariant operator, we will again use a scattering state. As before, this allows us to make contact with operators defined in the context of celestial holography and asymptotic symmetries.

The one-point detector observable in pure Yang-Mills theory measures the distribution of color flux radiated in a given direction. 
In the celestial representation, the limit $\Delta \to 2$ isolates the contribution associated with the number of particles. 
As before, we have the one-point function for a general GEO,
\begin{equation}
\mathbb{E}_{\Delta}\equiv \langle m | \mE_\Delta | m \rangle 
= \frac{1}{N_m}
 \SumInt_{k} \int_0^\infty d\omega\;
  \omega^{\Delta - 1}\,
  (2\pi)^4\delta^{(4)}(\omega q + K - P)
  \Big| \mathcal{M}^{a_1\cdots a_m\,a_{m+1} \cdots a_{m+n+1}}_{m \to n+1} \Big|^2 .
\label{eq:detector_def}
\end{equation}
The states $| m \rangle$ denote
$$
| m \rangle= | (p_1,\sigma_1,a_1)\cdots(p_m,\sigma_m,a_m)\rangle\,,
$$
an $m-$particle state with helicities $\sigma_k$, color denoted by $a_k$ and total momentum $P$. Unless necessary, we omit color and helicity index from the states.  For clarity, the amplitude $\mathcal{M}^{a_1\cdots a_m\,a_{m} \cdots a_{m+n}}_{m \to n}$ carries adjoint color indices on each external gluon.

As in the gravity case, the $\Delta\to 2$ limit is controlled by interactions of soft quanta. Therefore, the amplitude with one additional soft emission is controlled by the soft-gluon theorem,
\begin{equation}
\lim_{\omega \to 0}
\mathcal{M}^{a_1\cdots a_m\,a_{m+1} \cdots a_{m+n+1}}_{m \to n+1}
  = \frac{g_{\rm YM}}{\omega}
  \left[
  \sum_{k \in \text{out}} \frac{p_k \!\cdot\! \varepsilon}{p_k \!\cdot\! q}\, (T^{a}_k)_{a_k,b}
  - \sum_{k \in \text{in}} \frac{p_k \!\cdot\! \varepsilon}{p_k \!\cdot\! q}\, (T^{a}_k)_{a_k,b}
  \right]
\mathcal{M}^{a_1\cdots b \cdots a_{m+n}}_{m \to n} \,.
\label{eq:softthm}
\end{equation}
Here we have introduced the color operator $T^{a}_k$ whose action on gluons is 
\begin{equation}
(T^a_k)_{a_k, b}=if^{a b a_k}.
\end{equation}
Squaring the amplitude gives the soft limit of the probability density
\begin{equation}
|\mathcal{M}|^2_{\text{soft}}
  = \left(
     \sum_{i,j \in \text{in}} 
     - \sum_{i \in \text{in}, j \in \text{out}}\{\cdot,\cdot\}
       +  \sum_{i,j \in \text{out}} 
    \right)
     g_{\rm YM}^2 \,K^{(1)}(x_i,x_j,x)\, T_i \!\cdot\! T_j \,\,  \big| \mathcal{M}_{m \to n} \big|^2\,,
\label{eq:softM2}
\end{equation}
where repeated color indices are contracted with $\delta^{ab}$, and the middle term includes an anticommutator for the color operators.
We have suppressed explicit color indices on the amplitude for brevity, and we see again the appearance of the eikonal kernel \eqref{eq:eikonalkernel}.
Using manipulations similar to those in the previous subsection, the one-point detector becomes
\begin{equation}
\label{eq:celestial_final}
   \lim_{\Delta \to 2}
  (\Delta - 2)\,
  \mathbb{E}_\Delta
  =  g_{\rm YM}^2  \int d^2y_1\, d^2y_2\;
  K^{(1)}(y_1,y_2,x)\,
  \Big\langle
  m \Big|
    (N_{\text{out}}^{a}(y_1)\,
    -N_{\text{in}}^{a}(y_1))
    (N_{\text{out}}^{a}(y_2)\,
    -N_{\text{in}}^{a}(y_2))
  \Big| m \Big\rangle .
\end{equation}
where we have identified the color interference number detector for gluons as \cite{Chang:2025zib} 
\begin{equation}
N^a(y) = \frac{1}{(2\pi)^3}\int^\infty_0 d\omega \, \omega  \sum_{\sigma=\pm} if^{abc} a^{\dagger b}_\sigma(\omega,y) a^{c}_\sigma(\omega,y)\,.
\label{eq:density}
\end{equation}
The above operator acts on $m$-particle states as the sum  of the color operator $T^a_{k}$ acting on each particle~\cite{Gonzo:2019fai},
\begin{equation}
N^a(y)|m\rangle= \sum^{m}_{k=1}  \delta^{(2)} (y-z_k)T^a_k\,|m\rangle. 
\end{equation}
Equation~\eqref{eq:celestial_final} expresses the one-point detector observable as a bilocal correlator of in/out color fluxes, $N_{\text{in/out}}^a(y)$, weighted by the kernel $K^{(1)}$. 

As in gravity, a classical phase space analysis of Yang--Mills shows that the theory is invariant under an infinite-dimensional symmetry algebra given by color charge for every point of the sphere \cite{Strominger:2013lka,He:2015zea,He:2020ifr}. The corresponding conservation law reads
\begin{equation}
2\pi i (N^a_{\rm out}-N^a_{\rm in})= \partial_z J^a_{\bar{z}}+\partial_{\bar{z}} J^a_{z}\,,
 \label{HSmatching}
\end{equation}
where we have introduced the soft current defined in \cite{He:2015zea}
\begin{equation}
J_i\equiv  \frac{4\pi}{g^2_{\rm YM}}\left( \int^\infty_{-\infty} du F^{(0)}_{ui}-\int^\infty_{-\infty} dv F^{(0)}_{vi}  \right)\,.
\end{equation}
Here, $F^{(0)}_{ui}$ and $F^{(0)}_{vi}$ are the leading field strength components close to future and past null infinity, respectively.

Using this conservation law, we can rewrite
\begin{equation}
    \begin{split}
     \lim_{\Delta \to 2}
  (\Delta - 2)\,
  \mathbb{E}_\Delta(x)
  =  -\frac{g^2_{\rm YM}}{4\pi^2}\int d^2y_1\, &d^2y_2\;
  K^{(1)}(y_1,y_2,x)
 \\ \times &\langle
  m \Big|\Big(\partial_{\zbar}J^a_{z}(y_1)+\partial_{z}J^a_{\zbar}(y_1)\Big)\Big(\partial_{\zbar}J^a_{z}(y_2)+\partial_{z}J^a_{\zbar}(y_2)\Big)
  \Big| m \Big\rangle .
  \end{split}
\end{equation} 
This result can be simplified using the distributional identities
\begin{equation}
\begin{split}
\label{eq:gaugeidentity}
\bar{\partial}_1 \partial_2K^{(1)}(x_1,x_2,x)&=-\pi^2 \delta^{(2)}(z-z_1)\delta^{(2)}(z-z_2)-\frac{1}{(\bar{z}-\bar{z}_1)^2(z-z_2)^2}\,,\\
\partial_1 \partial_2K^{(1)}(x_1,x_2,x)&=\frac{\pi}{(z-z_1)^2} \delta^{(2)}(z-z_2)+\frac{\pi}{(z-z_2)^2} \delta^{(2)}(z-z_1) \,,
\end{split}
\end{equation}
and analogous expression for $\bar{\partial}_1 \bar{\partial}_2K^{(1)}$ obtained by complex conjugation. As before, the rational contributions yield the shadow transforms of the soft currents. We therefore find the final result
\begin{equation}
   \lim_{\Delta \to 2}
  (\Delta - 2)\,
  \mathbb{E}_\Delta
  =\tfrac{1}{2}g_{\rm YM}^2  \Big\langle
  m \Big|(-\widetilde{J^a_{\zbar}}+J^a_z)(-\widetilde{J^a_{z}}+J^a_{\zbar}) \Big| m \Big\rangle.
\end{equation}
 We have chosen to keep the shadow transforms explicit, as the precise definition of the associated symmetry current in non-Abelian gauge theories remains subtle and is not essential for the present analysis~\cite{He:2015zea,Himwich:2025bza}. Again, we obtain a bilinear expression in direct analogy with the structure found in gravity. In particular, the soft energy operator appears as the natural composite of two conformally soft currents.

We can once more perform an independent check of this bilinear formula in the self-dual Yang-Mills theory. In this case, the colored celestial OPE \cite{Pate:2019lpp} for gluons of definite helicity completely determines the product expansion $\mathcal{N}\mathcal{E}_{\Delta}$. Using this OPE, one directly reproduces \eqref{OPE:N} for $s=1$.

Finally, we note that, in contrast to gravity, the leading soft-gluon theorem \eqref{eq:softthm} receives loop corrections \cite{Catani:2000pi}. We can therefore expect that the above relation will also receive corrections at the loop level, which implies an IR renormalization of the soft currents. 

\section{Concluding remarks}
\label{sec:conclusion}
We have combined the study of detector operators with recent insights into asymptotic symmetries and the flat space holography program in general, initiating a reorganization of infrared physics in massless gauge theory and gravity through the language of detectors. 

We showed that both the Carrollian and the celestial formulation of massless scattering are useful in this regard: the former to construct detector operators using conformal Carrollian primary fields that are tailored to the structure of null infinity; the latter to examine the soft properties of detector operators.

We focused on two universal sectors of detector operators. The first is governed by collinear physics: at leading order, the detector OPE is completely fixed by the universal splitting functions, whose structure determines how scalar and spinning detectors combine. The second emerges from symmetry considerations. In fact, the pole structure of the OPE singles out the 
$\Delta=2$ detector, which plays a distinguished role as the operator counting the number of outgoing particles.

While this detector itself is not finite, its pole can be realized as a bilinear of soft currents. This means that particle number is not an independent observable, but one that is fixed entirely by the symmetries acting at $\mathscr{I}^{+}$. One can regard this as another manifestation of the relation between the gravitational memory effect and BMS transformations \cite{Strominger:2014pwa}.
More speculatively, it is also reminiscent of a Sugawara-type construction, where composite operators are determined by the currents that generate the symmetry algebra (for application to celestial holography see \cite{Fan:2020xjj}). This suggests that the soft sector may admit a decoupled CFT organization (see \cite{Kapec:2021eug, Nguyen:2021ydb, Gonzalez:2021dxw, Magnea:2021fvy,Donnay:2023kvm} for related ideas), with detector operators at special values of $\Delta$ serving as natural building blocks. 

The observation made in  \cite{Chang:2025zib} that detector Regge trajectories intersect precisely at the points where soft theorems appear hints at a universal “soft–detector” algebra. In our formulation, these intersections naturally correspond to the composite nature of detector operators in terms of soft currents, suggesting that the IR sector of gauge theory and gravity organizes itself through the algebra of detectors at $\mathscr{I}^{+}$. 

There are several directions for further investigation.
\begin{itemize}
    \item The points  $\Delta - 2 \in \mathbb{Z}$ at which the detector OPE coefficients develop poles coincide with the subleading soft limits \cite{Cachazo:2014fwa,Campiglia:2016efb, Strominger:2021mtt,Briceno:2025ivl}, showing that the analytic structure of the detector OPE directly encodes the hierarchy of soft theorems. Indeed, one can schematically see this by assuming a soft factorization of the form  $\mathcal{M}_{N+1}=\left[\sum_{n} \omega^n s_{(n)}\right] \mathcal{M}_N$. Then, the one-point function is 
    $$\langle \mathcal{E}_{\Delta} \rangle \propto \sum_{m,k} \int^\delta_0 d\omega\,  \omega^{\Delta+k-1} s_{(k-m)}s^{*}_{(m)} \,|\mathcal{M}_N|^2\,,$$ 
where we have introduced the soft cutoff $\delta$. Integrating close to this region, we clearly exhibit the conformally soft poles at  $\Delta=2,1,0,-1\dots$ as mentioned above. In particular, we expect that the operator $\Delta=1$ is related to the infinite-dimensional Virasoro symmetry coming from the superrotation charge.
\item 
Another obvious extension is to explore other detectors such as those sensitive to the angular momentum of the outgoing radiation or to the wave form of the radiation~\cite{Korchemsky:2021okt}.  They have received less attention in the literature but can be naturally constructed using the Carrollian language employed in Section~\ref{sec:detectorsdef}. It will be interesting to study their pole structure and how they encode polarization effects.
Ideally, this path can suggest a framework for constructing new gravitational wave observables.
\item 
The soft theorem dominates two-point functions in the back-to-back limit $(z_{ij}\to \infty)$~\cite{Chang:2025zib,Herrmann:2024yai}. This suggests that we can use the tower of leading and subleading symmetries to constrain two-point and higher-point functions at large Euclidean distance. 
\item 
We have shown that GEOs at specific weights are determined by symmetries, which implies a specific transformation behavior under large gauge transformations. In \cite{Gonzo:2020xza}, it was shown that the energy operator commutes with the soft charge, $[\mathcal{E}_3,Q_{\rm soft}]=0$. They suggested that this can be seen as showing IR-finiteness of the operator. It would be interesting to establish this as a definite criterion and apply it to the other GEOs and more general detector operators.
\end{itemize}
In this work, we took the first steps to show that detector operators provide a direct realization of asymptotic symmetries. Therefore, they are not only compatible with the celestial holography program,
but it is a natural language for it. From a broader perspective, our results give a unified framework that connects collider physics, light-ray operators, and the operator algebra of celestial fields at null infinity.

\acknowledgments
We thank Sruthi Narayanan for insightful comments. H.G. thanks Matías Briceño and Francisco Rojas for illuminating conversations. J.S. thanks Kévin Nguyen for many discussions on related topics. The authors thank the organizers of the workshop \emph{From Asymptotic Symmetries to Flat Holography: Theoretical Aspects and Observable Consequences} at the Galileo Galilei Institute for Theoretical Physics where parts of this work were performed.
The work of H.G. is funded by FONDECYT grant 1230853. J.S. is supported by a Postdoctoral Research Fellowship granted by the F.R.S.-FNRS (Belgium). 

\appendix
\section{Conventions}
\label{sec:conventions}
In the following we will give a summary of our conventions and other useful relations. We will  follow~\cite{Donnay:2022wvx}.
\subsection{Coordinates and polarization tensors}
The metric of $(d+2)$-dimensional Minkowski space is given by
\begin{equation}
    \dd s^2=\eta_{\mu\nu} \dd X^{\mu}\dd X^\nu,
\end{equation}
with $\eta_{\mu\nu}$ mostly plus.
Null vectors are parametrized by
\begin{equation}
  \label{eq:mompar}
  p^\mu(\omega,x^i)=\omega q^\mu(x^i),
\end{equation}
where $q^\mu(x^i)$ is the reference null direction
\begin{equation}
  \label{eq:qpar}
  q^\mu(x^i)=\frac{1}{\sqrt{2}}\left(1+x^2,2x^i,1-x^2\right), \qquad q^\mu(x^i)\cdot q^\mu(x^j)=-|x_i-x_j|^2 \qquad i=1,\ldots d,
\end{equation}
and $x^i$ is a cartesian coordinate on the $d$-dimensional plane.
The null vector
\begin{equation}
  \label{eq:npar}
  n^\mu=\frac{1}{\sqrt{2}}\left(1,0,-1\right),
\end{equation}
obeys $q \cdot n=-1$ and can be used to write Minkowski space in the coordinates
\begin{equation}
  \label{eq:6}
  X^\mu=u n^\mu+r q^\mu(x^i)
\end{equation}
with metric
\begin{equation}
  \label{eq:45}
  \dd s^2=-2\dd u\dd r+2r^2\delta_{ij}\dd x^i\dd x^j.
\end{equation}
The celestial sphere is therefore conformally mapped to a plane.

For the case of $d=2$, that is of main interest for this paper, we will often employ complex coordinates $z=x^1+i x^2$ in which the metric reads
\begin{equation}
  \dd s^2=-2\dd u\dd r+2r^2\dd z \dd \zbar.  
\end{equation}
We note that we take our measure as $d^2x=\dd x_1\dd x_2$, even if the integrand is written in terms of complex coordinates. 

We introduce the polarization vectors
\begin{equation}
    \veps^\mu_{\pm}=\frac{1}{\sqrt{2}}(\veps^\mu_1\mp i\veps^\mu_2)=\frac{1}{2}(\partial_1 q^\mu\mp i\partial_2 q^\mu).
\end{equation}
In complex coordinates, this takes the simple form $\veps^\mu_+=\partial_z q^\mu$ and its complex conjugate $\veps^\mu_-=\partial_{\bar{z}}q^\mu$. 
The projector onto the physical polarizations is given by
\begin{equation}
  \label{eq:vecprojector}
\Pi^{\mu\nu}=\veps^\mu_+ \veps^\nu_-+\veps^\mu_- \veps^\nu_+.
\end{equation}

Similarly, we define the polarization tensors for spin two field
\begin{equation}
    \veps^{\mu\nu}_{\pm}=\veps^{\mu\nu}_{11}\mp\veps^{\mu\nu}_{12}=\veps^{\mu}_\pm\veps^{\nu}_\pm.
\end{equation}
The polarization sum for spin two fields is given by
\begin{equation}
\label{eq:gravitonsum}
    \Pi^{\mu\nu\alpha\beta}=\sum_{i,j}\veps^{\mu\nu}_{ij}\veps^{\alpha\beta}_{ij}=\frac{1}{2}\left(\Pi^{\mu\alpha} \Pi^{\nu\beta}+\Pi^{\mu\beta} \Pi^{\nu\alpha}-\Pi^{\mu\nu}\Pi^{\alpha\beta}\right).
\end{equation}
\subsection{From bulk fields to boundary fields}
\label{sec:bulktobdy}
The canonical commutation relations for ladder operators of momentum $p_i=\omega_i q^\mu(x_i)$ are given by
\begin{equation}
  \label{eq:commrel}
  [a_I(p_1), a^\dagger_J(p_2)]=(2\pi)^32p^0_1\delta^{3}(p_1-p_2) \delta_{IJ}=(2\pi)^{3}\omega^{-1}\delta(\omega_1-\omega_2)\delta^{(2)}(x^i_1-x^i_2)\delta_{IJ},
\end{equation}
with additional Kronecker deltas for colored particles.
Accordingly, the free field expansion of a spin-s field in four-dimensional Minkowski space in De Donder gauge is given by
\begin{equation}
\begin{split}
  \label{eq:freefield}
  \Phi^{(s)}_{M}(X)&=K_{(s)}\sum_{J}\int\frac{d^{3}p}{(2\pi)^{3}2p^0}\left((\veps^J_M(p))^*a^{(s)}_J(p) e^{i p\cdot  X}+\veps^J_M(p)(a^{(s)}_J(p))^\dagger e^{-i p\cdot  X}\right)\\
  &=\frac{1}{(2\pi)^{3}}\sum_{J}\int^{\infty}_0\dd \omega\,\omega\int\dd^2 y\, \left((\veps^J_M)^*a^{(s)}_J e^{i \omega q(y)\cdot  X}+\veps^J_M(a^{(s)}_J)^\dagger e^{-i \omega q(y)\cdot  X}\right)
  \end{split}
\end{equation}
where $M$ denotes the multiindex $M=\{\mu_1\ldots \mu_s\}$ and $J$ denotes a multiindex for the polarization of the field. For gauge fields this is just $J=\{i\}$, while for gravitons one has $J=\{ij\}$.
The constant $K_{(s)}$ depends on the coupling constant, e.g., $K_{(1)}=e$ for a photon or $K_{(2)}=\sqrt{32\pi G\hbar}$ for a graviton.
Using the relation
\begin{equation}
  \label{eq:24}
 \lim_{r\to \infty}\int \dd^{d}y f(y) e^{\pm i \omega r |x-y|^2}\approx \frac{e^{\pm  i\pi d/4}}{2^{d}}\left(\frac{\pi}{\omega r}\right)^{\frac{d}{2}} f(x)
\end{equation}
and switching to complex coordinates we can extract the boundary field as
\begin{equation}
   \phi^{(s)}_{z\ldots z}(u,x^i)\dd z \otimes\ldots \dd z +\phi^{(s)}_{\zbar\ldots \zbar}(u,x^i)\dd \zbar \otimes\ldots \dd \zbar=\lim_{r\to \infty}(r^{1-s}\Phi^{(s)}_{\mu_1\ldots \mu_s}\dd X^{\mu_1}\ldots \dd X^{\mu_s})|_{r=const}.
\end{equation}
One finds the boundary operator
\begin{equation}
  \label{eq:boundarybulk}
  \phi^{(s)}_{z\ldots z}(u,x^i)=-i\frac{K_{(s)}}{8\pi^2}\int^\infty_0\dd \omega \left(a^{(s)}_{+}(\omega,x^i) e^{-i \omega u}+(a^{(s)}_{-}(\omega,x^i))^\dagger e^{i \omega u}\right)
\end{equation}
and its complex conjugate.

Under Poincaré transformations, this field transforms as
\begin{equation}
\label{Carrollian induced rep}
\begin{split}
\left[H, \phi^{(s)}_{\Delta}(u,x^i)\right]&=i \partial_u \phi^{(s)}_{\Delta}(u,x^i)\,,\\
\left[P_i, \phi^{(s)}_{\Delta}(u,x^i)\right]&=i \partial_i \phi^{(s)}_{\Delta}(u,x^i)\,,\\
\left[J_{ij}, \phi^{(s)}_{\Delta}(u,x^i)\right]&=i(-x_i \partial_j+x_j\partial_i -i\Sigma^{(s)}_{ij})\,\phi^{(s)}_{\Delta}(u,x^i)\,,\\
\left[D, \phi^{(s)}_{\Delta}(u,x^i)\right]&=i(\Delta+u\partial_u+x^i\partial_i )\,\phi^{(s)}_{\Delta}(u,x^i)\,,\\
\left[K, \phi^{(s)}_{\Delta}(u,x^i)\right]&=i x^2 \partial_u \phi^{(s)}_{\Delta}(u,x^i)\,,\\
\left[K_i, \phi^{(s)}_{\Delta}(u,x^i)\right]&=i(-2x_i \Delta-2ix^j \Sigma^{(s)}_{ij}-2x_i u\partial_u-2x_i x^j \partial_j+x^j x_j \partial_i )\,\phi^{(s)}_{\Delta}(u,x^i)\,,\\
\left[B_i, \phi^{(s)}_{\Delta}(u,x^i)\right]&=ix_i\partial_u \phi^{(s)}_{\Delta}(u,x^i)\,,
\end{split}
\end{equation}
with $\Delta=1$ (in general $\Delta=\frac{d}{2})$
where $\Sigma^{(s)}_{ij}$ denotes the spin $s$ representation of $\mathfrak{so}(2)$ or, more generally, $\mathfrak{so}(d)$. In the above, we used a different basis for the Poincaré algebra that arises naturally from regarding it as the conformal algebra of a Carrollian structure in one dimension lower.\footnote{This should be compared to the basis change between viewing $\mathfrak{so}(3,2)$ as the symmetries of $\textrm{AdS}_4$ or as the conformal symmetries of three-dimensional Minkowski space. Similar, to the case at hand the difference lies in their structure as homogeneous spaces \cite{Figueroa-OFarrill:2021sxz}.}
In the above basis, Poincaré translations correspond to the set $\{H,K_i,K\}$ while the remaining generators correspond to the Lorentz algebra written as the Euclidean conformal algebra in two dimension; see~\cite{Banerjee:2018gce,Nguyen:2023vfz,Donnay:2022wvx} for more details and the explicit basis changes. This transformation behavior defines a conformal Carrollian primary field. A time-independent conformal Carrollian field $\partial_u\phi^{(s)}_{\Delta}=0$ thus transforms as a primary field of the Euclidean conformal algebra $\mathfrak{so}(3,1)$.
\bibliography{bibl}
\bibliographystyle{jhep}
\end{document}